\documentclass[journal]{IEEEtran}

\ifCLASSINFOpdf
 \else
\fi

\usepackage{wrapfig}
\usepackage{comment}
\usepackage{balance}  
\usepackage{cite}
\usepackage{graphicx}
\usepackage{float}
\usepackage{algorithm}
\usepackage{algpseudocode}
\usepackage{amsmath} 
\usepackage[utf8]{inputenc}
\usepackage[final]{pdfpages} 
\usepackage{pdfpages}
\usepackage{lipsum}
\usepackage{textcase}
\usepackage{url}
\usepackage{amsmath,esint}
\usepackage{epstopdf}
\usepackage{array} 
\usepackage{hyperref}
\usepackage{multirow}
\usepackage{booktabs}
\usepackage{esvect}
\usepackage{soul}
\usepackage{ulem}

\newcolumntype{P}[1]{>{\centering\arraybackslash}p{#1}}
\newcolumntype{M}[1]{>{\centering\arraybackslash}m{#1}}

\pdfoutput=1

\begin{document}
%

\title{Channel Prediction for mmWave Ground-to-Air Propagation under Blockage}

\author{
\IEEEauthorblockN{Wahab Khawaja, Ozgur Ozdemir, \IEEEmembership{Member, IEEE}, and 
Ismail Guvenc} \IEEEmembership{Fellow, IEEE}\\
\thanks{This work was supported by NASA under the award NNX17AJ94A. W. Khawaja is 
with the Mirpur University of Science \& Technology, Pakistan. The co-authors are with NC State University, Raleigh, NC 27606 USA (e-mails: wahab.ali@must.edu.pk, \{wkhawaj, oozdemi, iguvenc\}@ncsu.edu). }
}


\maketitle

\begin{abstract}
Ground-to-air~(GA) communication using unmanned aerial vehicles~(UAVs) has gained popularity in recent years and is expected to be part of 5G networks and beyond. However, the GA links are susceptible to frequent blockages at millimeter wave (mmWave) frequencies. During a link blockage, the channel information cannot be obtained reliably. In this work, we provide a novel method of channel prediction during the GA link blockage at $28$~GHz. In our approach, the multipath components~(MPCs) along a UAV flight trajectory are arranged into independent path bins based on the minimum Euclidean distance among the channel parameters of the MPCs. After the arrangement, the channel parameters of the MPCs in individual path bins are forecasted during the blockage. An autoregressive model is used for forecasting. The results obtained from ray tracing simulations indicate a close match between the actual and the predicted mmWave channel.
\end{abstract}


\IEEEpeerreviewmaketitle

\section{Introduction}

The use of civilian unmanned aerial vehicles~(UAVs) for everyday applications have seen a surge in recent years such as surveillance, video recording, search and rescue, and hot spot communications. However, the UAV ground-to-air~(GA) communication links are susceptible to blockages due to high rise buildings or trees. These blockages are significant at millimeter wave~(mmWave) frequencies compared to sub-6~GHz frequencies. Channel prediction can be used to predict the state of the channel during blockages and improve link reliability, which has not been studied in the literature for GA scenarios to our best knowledge.


The literature for channel prediction can be divided into three main categories. 1)~Autoregressive~(AR) model based methods~\cite{Survey_Hallen,AR2}, 2)~Sum of sinusoids~(SoS) \cite{SoS1,SoS2}, and basis expansion models~(BEM)~\cite{BEM1,BEM2}, and 3)~Artificial intelligence~(AI) based models~\cite{NN1,NN2,NN3}. AR based prediction methods are the most popular, frequently used, and computationally simple, but their accuracy suffers in fast and complicated varying channels. The SoS methods require rigorous channel prediction calculations over short time durations for fast varying channels that become computationally not feasible for real-time environments and are therefore mostly used for simulated environments. The computational complexity and accuracy for the BEM are dependent on the type of the basis function and 
the type of the channel. 
The AI based  recurrent neural networks~(RNN) have gained popularity in recent years for channel prediction~\cite{NN1}, which have a high computational load for the training at regular intervals, as they require large memory and data for fast varying channels and complex environments. 
Channel prediction aided with the 3D maps of the urban propagation environment~\cite{3Dmap,3D_map2}, are also available in the literature. For example, in \cite{UAV1,UAV2}, the prediction of signal strength for trajectory planning of relay UAVs is aided with the help of maps. Comprehensive 3D data is required for the whole environment for map aided channel prediction methods. Also, the prediction accuracy suffers if the maps are outdated and real-time environmental variations are not included in the maps.

\begin{figure}[!t]
	\centering\vspace{-2mm}
	\includegraphics[width=\columnwidth]{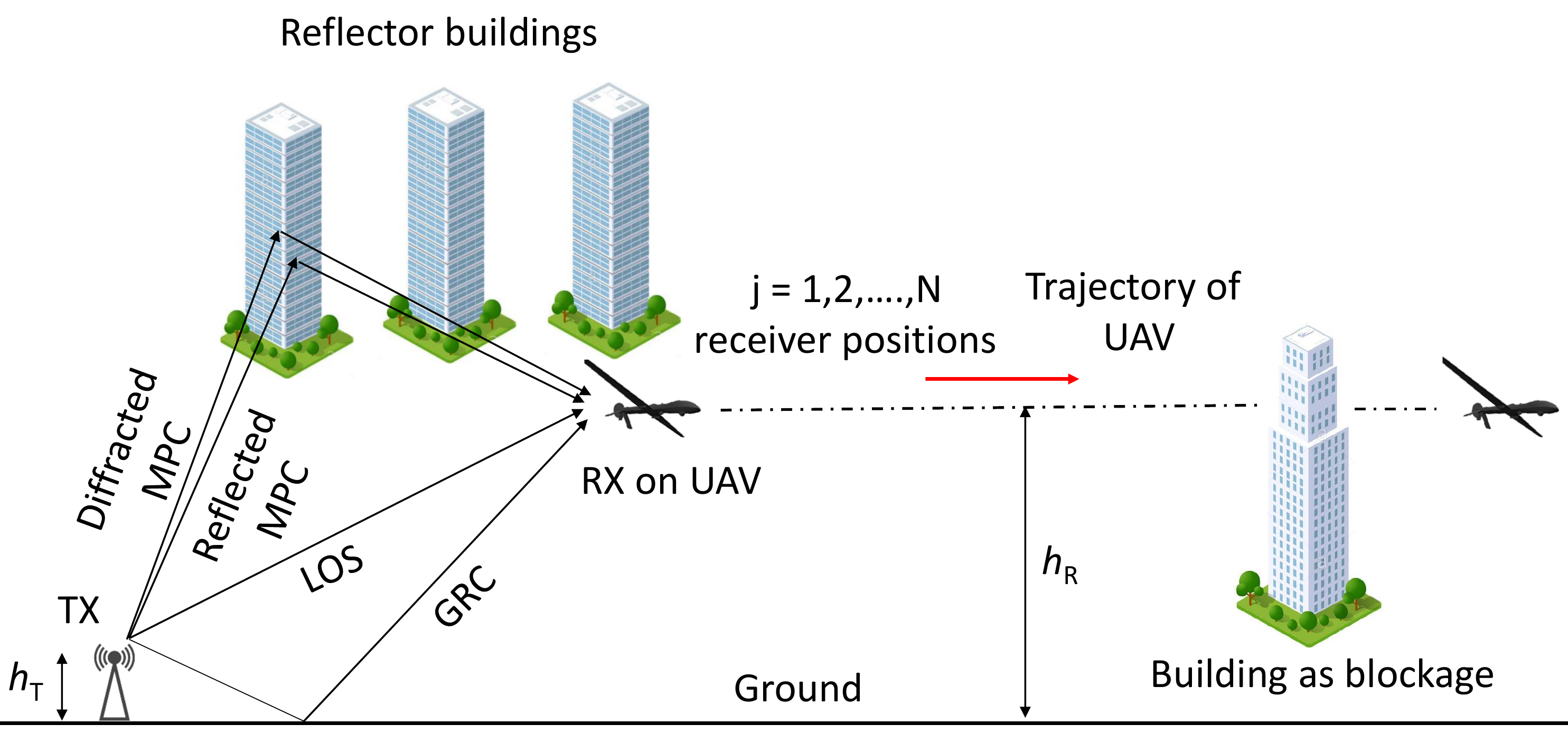}\vspace{-3mm}
	\caption{The UAV GA propagation scenario with blockage. }\label{Fig:LOS_Refl_diff}\vspace{-5mm}
\end{figure}

In this work, we have introduced a novel channel prediction method for GA communication links. The GA propagation scenario for our work is shown in Fig.~\ref{Fig:LOS_Refl_diff}. Our approach can be divided into two parts. In the first part, an Euclidean distance~(ED) based algorithm is used, which arranges the MPCs obtained along the UAV trajectory into individual path bins. The arrangement is based on a minimum ED among the channel parameters of the MPCs, and it results in a trend of individual channel parameters of MPCs in a path bin. In the second part, the trend of individual channel parameters in a path bin is used to forecast the channel parameters during the blockage. The forecasting is performed using an auto-regressive~(AR) model. The simulations are carried out using the  Wireless InSite ray tracing software. The results indicate that the actual and the forecasted channel parameters are close. 
We also provide a distance-based prediction approach for the death of a path bin along the UAV trajectory, and performance comparison with other popular prediction methods.


\section{System Model} We consider a GA propagation scenario shown in Fig.~\ref{Fig:LOS_Refl_diff}. The transmitter~(TX) is on the ground and the receiver~(RX) is on a UAV. The height of the TX and RX above the ground is $h_{\rm T}$ and $h_{\rm R}$, respectively. The received MPCs in Fig.~\ref{Fig:LOS_Refl_diff} are categorized as: i) line-of-sight~(LOS), ii) ground reflected component~(GRC), and iii) reflected and/or diffracted from scatterers. We consider six channel parameters of each MPC: two in the temporal and four in the spatial domain. 


\subsection{Euclidean Distance of MPCs' Channel Parameters} \label{Section:Mathematical_model} 
Let $H(n)$ represent the time-varying channel impulse response of the system. Then,  we may write
\small
\begin{align}
    &\big|H(n)\big|^2=\sum_{m=1}^{M}\big|\alpha_m (n)\big|^2\delta\big(n-\tau_m (n)\big) \delta\big(\vv{\boldsymbol{\Theta}}^{\rm (TX)}-\theta_m^{\rm (TX)}\big) \nonumber \\
& \delta\big(\vv{\boldsymbol{\Phi}}^{\rm (TX)}-\phi_m^{\rm (TX)}\big) \delta\big(\vv{\boldsymbol{\Theta}}^{\rm (RX)}-\theta_m^{\rm (RX)}\big) \delta\big(\vv{\boldsymbol{\Phi}}^{\rm (RX)}-\phi_m^{\rm (RX)}\big), \label{Eq:CIR_tavg}
\end{align}
\normalsize
where $n$ is the time instance, $M$ is the total number of MPCs, $\alpha_m (n)$, $\tau_m (n)$ represents the complex amplitude and delay of the $m^{\rm th}$ MPC, respectively, $\vv{\boldsymbol{\Theta}}^{(\rm TX)}, \vv{\boldsymbol{\Phi}}^{(\rm TX)}$ are the angle of departure~(AoD) vectors in the elevation and azimuth planes, respectively, $\vv{\boldsymbol{\Theta}}^{(\rm RX)}, \vv{\boldsymbol{\Phi}}^{(\rm RX)}$ are AoA vectors in the elevation and azimuth planes, respectively, and the AoD and AoA of the $m^{\rm th}$ MPC in the elevation and azimuth planes respectively, are represented as $\theta_m^{\rm (TX)}$, $\phi_m^{\rm (TX)}$, $\theta_m^{\rm (RX)}$, $\phi_m^{\rm (RX)}$.

From (\ref{Eq:CIR_tavg}), if $P_{\rm T}$ is the transmitted power, the total received power $P_{\rm R}$ at a RX position is given as
\small
\begin{align}
P_{\rm R} &= \frac{P_{\rm T}\lambda^2}{(4\pi)^2}\bigg[\frac{G^{(\rm TX)}\big(\theta_1, \phi_1\big) G^{(\rm RX)}\big(\theta_1, \phi_1\big)}{R_1^2} +  \label{Eq:RX_pwr} \\ \nonumber &\frac{\sum_{s=0}^S \sum_{m=1}^{M_s} G^{(\rm TX)}\big(\theta_{s,m}, \phi_{s,m}\big) G^{(\rm RX)}\big(\theta_{s,m}, \phi_{s,m}\big)\Gamma_{s,m}}{R_{s,m}^2} \bigg],
\end{align}
\normalsize
where the first term is for the LOS and the second term represents the MPCs from scatterers, $\lambda$ is the wavelength, $G^{(\rm TX)}\big(\theta, \phi\big)$ and $G^{(\rm RX)}\big(\theta, \phi\big)$ are the antenna gains at the TX and RX, respectively, at elevation angle $\theta$ and azimuth angle $\phi$, $\Gamma_{s,m}$ is the reflection coefficient, and $R_{s,m}$ is the total distance traveled by the $m^{\rm th}$ MPC due to interaction with the $s^{\rm th}$ scatterer, and the distance of the LOS path is represented as $R_1$. Moreover, the total number of scatterers are $S$ and the total number of MPCs due to a scatterer is represented as $M_s$.     

\begin{algorithm}[!t]
\small
\caption{Arrangement of MPCs into path bins.}\label{Alg:Algorithm}
\begin{algorithmic}[1]
\Procedure{Pathbins}{}\\
Initialize each path bin with the MPC at first RX position \\
\% At the $j^{\rm th}$ RX position and from (\ref{Eq:distance1}) and (\ref{Eq:distance2}), the ED is
\For{$m=1:M_j$}~\% MPCs at $j^{\rm th}$ RX position
\For{$i=j-1:1$} ~\% previous 
\For{$k=1:M_i$} ~\% MPCs at previous 
\State Calculate $d(j,m,i,k)$ 
\EndFor
\EndFor
\State $d_{\rm min}(m)$ = $\displaystyle \min_{\forall i,k} d$
\If {$d_{\rm min}(m)<\epsilon$} 
\For{$l=1$:Number of path bins}
\State (i) Select min$(d_l)$, and, 
\State (ii) add the $m^{\rm th}$ MPC at $j^{\rm th}$ RX position to $l^{\rm th}$ 
\State path bin. \% MPCs placement in a path bin
\EndFor
\EndIf
\If {$d_{\rm min}(m)>\epsilon$} 
\State (i) Birth of a new MPC and a path bin, and, 
\State (ii) temporary discontinuation of a path bin
\EndIf
\EndFor
\If {$M_j<M_{j-1}$} 
\State (i) Death of a MPC, and,
\State (ii) temporary discontinuation of corresponding path bin 
\EndIf\\
\Return{path bins}
\EndProcedure
\end{algorithmic}
\end{algorithm}

From Fig.~\ref{Fig:LOS_Refl_diff}, the RX positions along the UAV trajectory are labeled as $j=1,2,...,N$. At a $j^{\rm th}$ RX position and $m^{\rm th}$ MPC, a MPC vector represented as $\vv{\textbf{MC}}_{j,m}$ is given by: 
\small
\begin{align}
\vv{\textbf{MC}}_{j,m} = \big[\alpha_{j,m}, \tau_{j,m}, \theta_{j,m}^{(\rm TX)}, \phi_{j,m}^{(\rm TX)}, \theta_{j,m}^{(\rm RX)}, \phi_{j,m}^{(\rm RX)}\big].\label{eq:sixchpar}
\end{align}
\normalsize
The MPCs till the $(j-1)^{\rm th}$ RX position are given by: 
\small
\begin{equation}
\begin{bmatrix}
\vv{\textbf{MC}}_{1,1}&\vv{\textbf{MC}}_{1,2}&\vv{\textbf{MC}}_{1,3}&\cdots & \vv{\textbf{MC}}_{1,M_1}\\
\vv{\textbf{MC}}_{2,1}&\vv{\textbf{MC}}_{2,2}&\vv{\textbf{MC}}_{2,3}&\cdots & \vv{\textbf{MC}}_{2,M_2}\\
\vdots & \vdots &\vdots &\cdots& \vdots\\
\vv{\textbf{MC}}_{j-1,1}&\vv{\textbf{MC}}_{j-1,2}&\vv{\textbf{MC}}_{j-1,3}&\cdots & \vv{\textbf{MC}}_{j-1,M_{j-1}}
\end{bmatrix}~,  \nonumber
\end{equation}
\normalsize
which is used to calculate the ED of a MPC at $j^{\rm th}$ RX position with the MPCs at RX positions $[j-1, j-2, \hdots ,1]$. The number of previous RX positions for ED calculation can be varied to reduce the channel data. The ED between the channel parameters of the $m^{\rm th}$ MPC at $j^{\rm th}$ RX position and the channel parameters of $k^{\rm th}$ MPC at $i^{\rm th}$ RX position, where $i= j-1, j-2, \hdots ,1$, is given by: 
\small
\begin{align}
d^{(v)}(j,m,i,k) = {\rm Euclidean} \big(\vv{\textbf{MC}}_{j,m}(v), \vv{\textbf{MC}}_{i,k}(v) \big) ~,
\label{Eq:distance1}  \end{align} \normalsize
where $v=1,2,...,6$ refers to the six channel parameters of a MPC as in~\eqref{eq:sixchpar}. The total ED 
of the $m^{\rm th}$ MPC at $j^{\rm th}$ RX position, from the $k^{\rm th}$ MPC at $i^{\rm th}$ RX position is given by
\small
\begin{align}
d(j,m,i,k) = \frac{1}{\gamma}\sum_{v=1}^6d^{(v)}(j,m,i,k)~, \label{Eq:distance2}
 \end{align}
 \normalsize
where $\gamma$ is the normalizing factor. The value of $\gamma$ can vary from $1$ to $\max\big(\sum_{v=1}^6d^{(v)}(j,m,i,k)\big) ~\forall ~j,m,i,k$.

\subsection{Channel Prediction Algorithms}
The pseudocode for arranging the MPCs in path bins is shown in Algorithm~\ref{Alg:Algorithm}, where two minimum ED conditions are used for a MPC to be placed in a path bin. The first condition takes the distance of the $m^{\rm th}$ MPC at a $j^{\rm th}$ RX position with the MPCs at previous RX positions. The distance is calculated among individual channel parameters of MPCs represented from $d^{(1)}$ to $d^{(6)}$ in (\ref{Eq:distance1}). The distance of individual channel parameters are added and normalized by $\gamma$ to get $d$ in (\ref{Eq:distance2}). In Algorithm~\ref{Alg:Algorithm}, the minimum value of $d$ for the $m^{\rm th}$ MPC is compared with the threshold $\epsilon$, given as $d_{\rm min}(m)<\epsilon$. If this condition is true, the MPC is selected and a second minimum distance condition is applied. In the second condition, min$(d_l)$ is used, where $d_l$ is the ED of selected MPC with the existing MPCs in the $l^{\rm th}$ path bin. This condition selects the $l^{\rm th}$ path bin, where the MPC will be placed. The minimum distance, $d_{\rm min}$ threshold, $\epsilon$ in Algorithm~\ref{Alg:Algorithm} is dependent on the normalizing factor $\gamma$. 

\begin{figure}[!t]
	\centering
	\vspace{-0.3cm}
	\includegraphics[width=0.92\columnwidth]{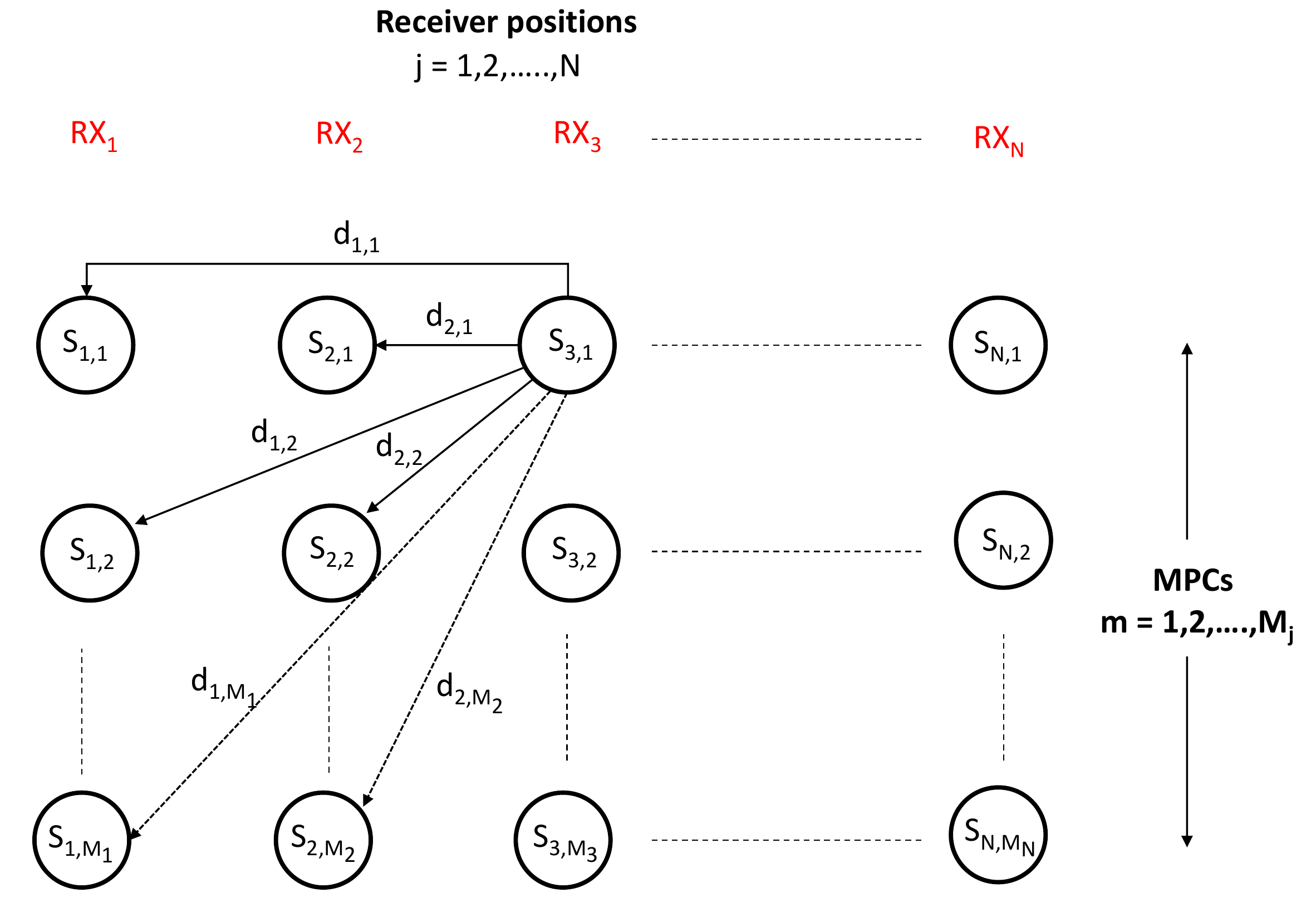}
	\caption{A Markov chain model based on the minimum ED.   }\label{Fig:Markov}
	\vspace{-0.1cm}
\end{figure} 

\begin{algorithm}[!t]
\small
\caption{Forecasting of the channel parameters in path bins during blockage.}\label{Alg:forecasting}
\begin{algorithmic}[1]
\Procedure{Forecasting}{path bins}\\
Let $\vv{\textbf{MC}}_{{pb}_1}$ represent the MPCs in the first path bin $\vv{\textbf{PB}}_1$, and $pb_1 = 1,2,...,M_{{pb}_1}$, where $M_{{pb}_1}$ is the total number of MPCs in $\vv{\textbf{PB}}_1$ over all the \\
The six channel parameters of the MPCs in $\vv{\textbf{PB}}_1$ from (\ref{eq:sixchpar}) are represented as $\vv{\textbf{MC}}_{{pb}_1}(v)$, ~$v=1,2,...,6$ \\
Fit an $n^{\rm th}$ order AR model to a channel parameter $\vv{\textbf{MC}}_{{pb}_1}(v)$ obtained over RX positions as
\State $S_{\rm AR}$ = AR\big($\vv{\textbf{MC}}_{{pb}_1}(v)$,~$n$\big)\\
If $K^{\rm (fc)}$ are the number of points to be forecasted, then the forecasted data $\vv{\textbf{MC}}_{{pb}_1}^{(\rm fc)}(v)$, based on system model $S_{\rm AR}$ and past data $\vv{\textbf{MC}}_{{pb}_1}(v)$ is
\State $\vv{\textbf{MC}}_{{pb}_1}^{(\rm fc)}(v)$ = $\rm {forecast}$\big($S_{\rm AR}$,~ $\vv{\textbf{MC}}_{{pb}_1}(v)$,~ $K^{\rm (fc)}\big)$\\
\Return{$\vv{\textbf{MC}}_{{pb}_1}^{(\rm fc)}(v)$}
\EndProcedure
\end{algorithmic}
\vspace{-1mm}
\end{algorithm}

From Algorithm~\ref{Alg:Algorithm}, if $d_{\rm min}(m)>\epsilon$, the similarity criteria for the $m^{\rm th}$ MPC with any of the MPCs at previous RX positions is not met. Therefore, this MPC is considered a new MPC~(birth), and a new path bin is created for it. Moreover, if the number of MPCs at $j^{\rm th}$ RX position is less than the $(j-1)^{\rm th}$ position, this indicates the death of a MPC. In both the above conditions, one of the existing path bins will be discontinued. This discontinuation can be temporary as these path bins can continue~(resurrect) at a later part of the UAV trajectory. The value of $\gamma$ in (\ref{Eq:distance2}) and $\epsilon$ in Algorithm~\ref{Alg:Algorithm} is $75.8$ and $0.15$, respectively. 

\begin{wrapfigure}{r}{0.5\columnwidth}
\vspace{-1mm}
	\centering
	\includegraphics[width=0.5\columnwidth]{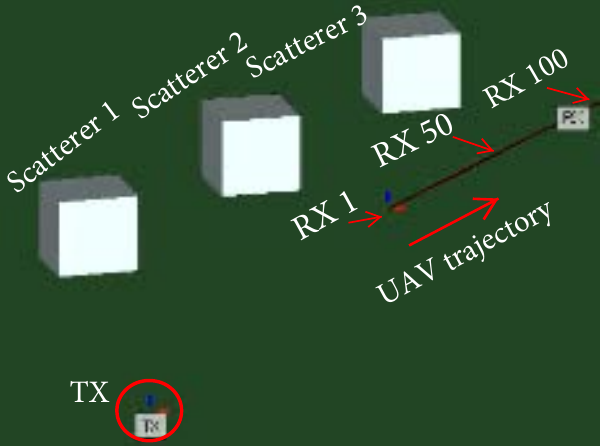}\vspace{-2mm}
	\caption{28~GHz simulation scenario in Wireless InSite ray tracing software.}\label{Fig:WI_scenario}\vspace{-1mm}
\end{wrapfigure}

From Algorithm~\ref{Alg:Algorithm}, the transition of MPCs along the RX positions based on ED in a path bin can be modeled using a Markov chain. A transition likelihood scenario based on the Markov chain for the first MPC at the third RX position is shown in Fig.~\ref{Fig:Markov}. The state of a Markov chain at $j^{\rm th}$ RX position and $m^{\rm th}$ MPC is represented as $S_{j,m}$. The transition among the states is dependent on the minimum ED. The minimum transition distance in Fig.~\ref{Fig:Markov} will determine which path bin the first MPC at the third RX position will occupy. Therefore, an $l^{\rm th}$ path bin from Algorithm~\ref{Alg:Algorithm} is represented as a sequence of selected Markov states along the RX positions. Algorithm~\ref{Alg:Algorithm} can handle any UAV trajectory. A change in the UAV trajectory is accompanied by the birth and death of MPCs that will be tracked and updated by Algorithm~\ref{Alg:Algorithm}.

\begin{table}[!t]\vspace{-2mm}
	\begin{center}
		\caption{Parameters for ray tracing simulations.}\label{Table:Simulations}\vspace{-2mm}
		\resizebox{0.9\columnwidth}{!}{
        \begin{tabular}{@{} |P{5.2cm}|P{2.5cm}| @{}}
			\hline
			\textbf{Parameter}&\textbf{Parameter value}\\			
			\hline
		    Center frequency& $28$~GHz \\
            \hline
            Antenna radiation pattern (azimuth)& Omnidirectional \\
            \hline
            Antenna polarization& Vertical \\
            \hline
            Height of TX, $h_{\rm T}$ & $2$~m \\
            \hline
            Height of RX~(on UAV), $h_{\rm R}$  & $50$~m \\
            \hline
            Length of UAV trajectory& $100$~m \\
            \hline
            Distance of TX to start of trajectory& $243$~m \\
            \hline
            Horizontal scatterer distance from trajectory & $145$~m \\
            \hline
            Dimension of scatterers 
            & 40~m$ \times $40~m$ \times$ 40~m \\
            \hline
            Distance among scatterers& $110$~m \\
            \hline
            Permittivity of ground
            &  $3.5$\\
            \hline
            Permittivity of scatterer structure
            & $5.31$ \\
            \hline
            \end{tabular}
            }
		\end{center}\vspace{-5mm}
\end{table}

 The placement of MPCs with similar channel characteristics in path bins results in a trend of individual channel parameters along the UAV trajectory. In case of a blockage, the trend of a channel parameter in a path bin is used to forecast it. Our approach for forecasting the channel parameters is given in Algorithm~\ref{Alg:forecasting}, where the individual channel parameters in a path bin form a time series. An $n^{\rm th}$ order AR model is fitted to individual channel parameters. The time series system model using autoregression is represented as $S_{\rm AR}$ in Algorithm~\ref{Alg:forecasting}. The forecasting over $K^{(\rm fc)}$ future steps is performed using a system model and past channel parameter data.

\section{Simulation Scenario and Results} \label{Section:Simulations}
\begin{figure}[!t]
	\centering\vspace{-1mm}
	\includegraphics[width=0.85\columnwidth, scale=0.5]{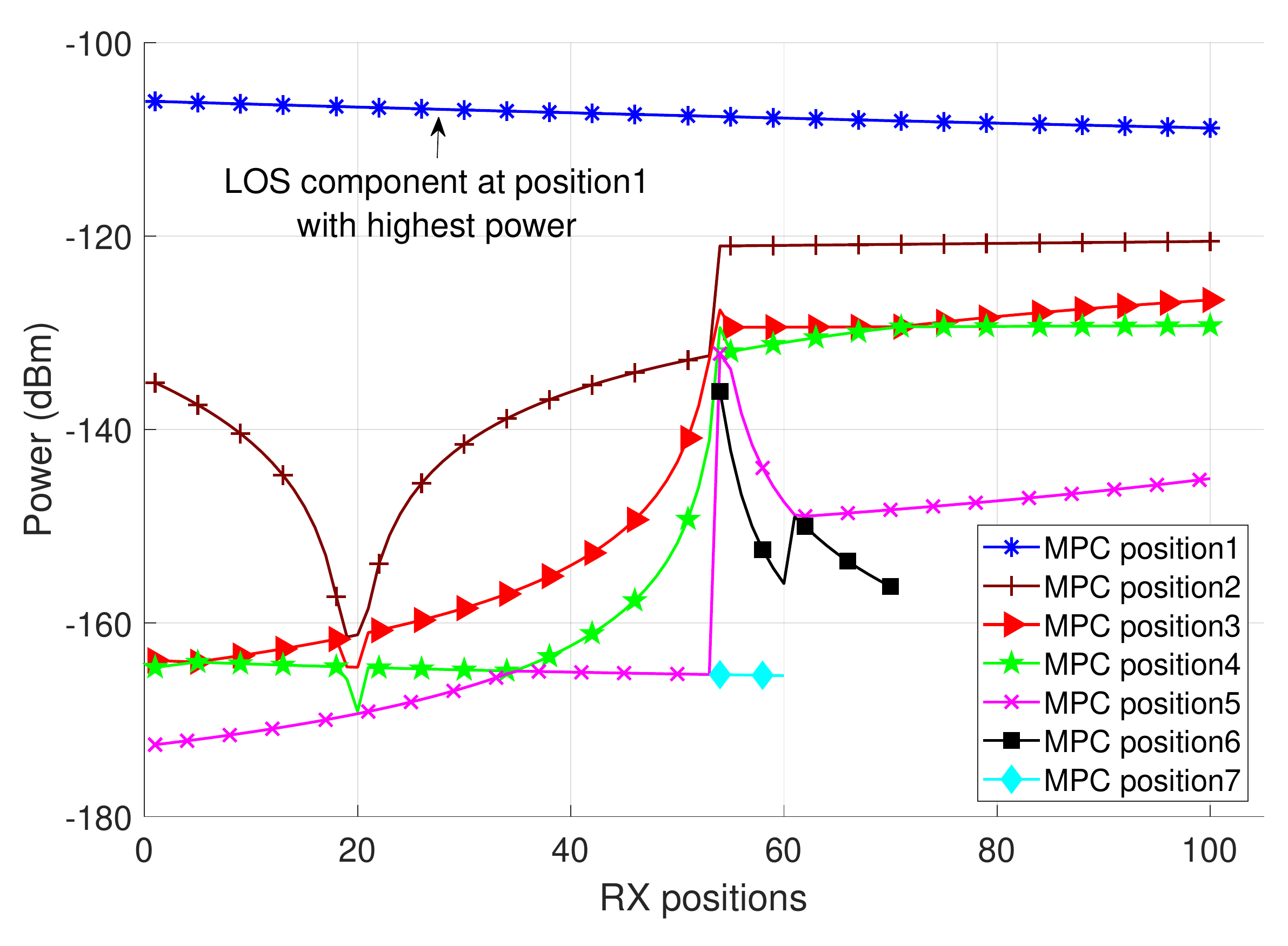}\vspace{-3mm}
	\caption{Power of received MPCs along the UAV trajectory without applying Algorithm~\ref{Alg:Algorithm}. The MPCs are positioned based on descending received power.}\label{Fig:raw_paths}\vspace{-2mm}
\end{figure}

The simulations are carried out using the Wireless InSite ray tracing software (see Fig.~\ref{Fig:WI_scenario} and Table~\ref{Table:Simulations}). Fig.~\ref{Fig:raw_paths} shows the received power of the MPCs along the RX positions. The MPCs are positioned based on descending received power at respective RX positions. For example, in the start of Fig.~\ref{Fig:raw_paths} at RX position~1, there are five MPCs. The LOS and GRC (with larger received power compared to other MPCs) occupy the first and second positions, respectively, followed by the other MPCs. However, from Fig.~\ref{Fig:raw_paths}, no clear pattern of received power of the MPCs over the RX positions can be observed, except for the LOS path. Also, we cannot differentiate the birth, and death of the MPCs along the RX positions. 

  \begin{figure}[!t]
	\centering\vspace{-2mm}
	\includegraphics[width=0.85\columnwidth, scale=0.5]{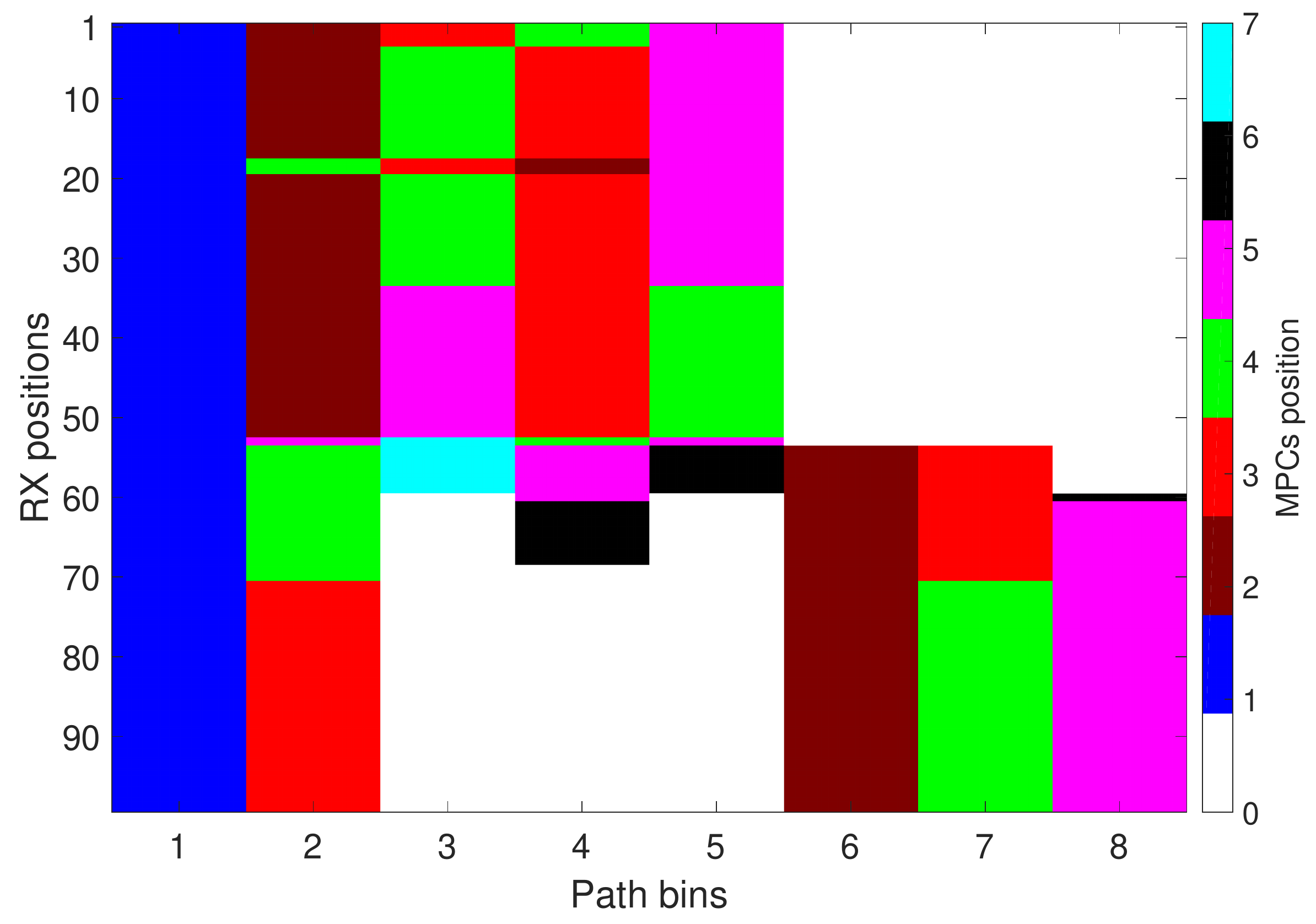}\vspace{-3mm}
	\caption{Position of the MPCs in individual path bins, sequenced using Algorithm~\ref{Alg:Algorithm}. For example, the LOS path is shown in path bin $1$. }\label{Fig:Path_bins}\vspace{-3mm}
\end{figure}

\subsection{MPC Identification and Channel Gain Prediction}

\begin{figure}[!t]
	\centering\vspace{-1mm}
	\includegraphics[width=0.85\columnwidth, scale=0.5]{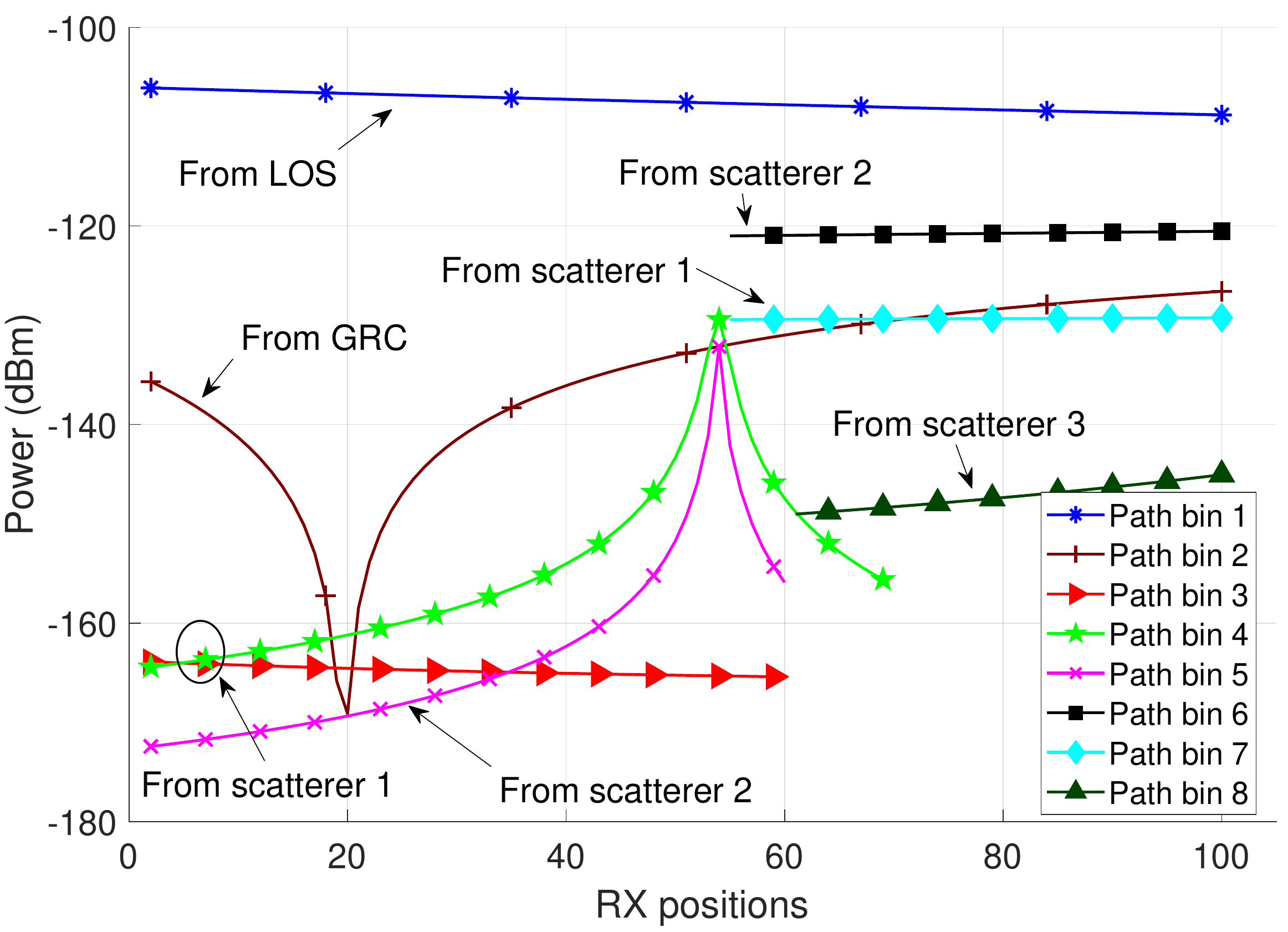}\vspace{-3mm}
	\caption{Received power of MPCs sequenced in individual path bins using Algorithm~\ref{Alg:Algorithm}. The LOS and source of the MPCs in path bins are also provided.}\label{Fig:separated_paths}\vspace{-1mm}
\end{figure}

The selected MPCs (or states in Fig.~\ref{Fig:Markov}) from Algorithm~\ref{Alg:Algorithm}, are placed in the respective path bins. The arrangement of the MPCs in path bins at RX positions is shown in Fig.~\ref{Fig:Path_bins}. Compared to Fig.~\ref{Fig:raw_paths}, the MPCs in Fig.~\ref{Fig:Path_bins} are placed in individual path bins based on the similarity of the channel parameters. Moreover, the number of path bins is controlled by the similarity criteria set by $\epsilon$ in Algorithm~\ref{Alg:Algorithm}. Fig.~\ref{Fig:separated_paths} shows the received power of the MPCs in path bins after using Algorithm~\ref{Alg:Algorithm}. A clear trend of the received power of the MPCs is observed in Fig.~\ref{Fig:separated_paths}, as opposed to the results in Fig.~\ref{Fig:raw_paths}. This trend is used for forecasting the received power during the blockage. The birth and death of the MPCs in Fig.~\ref{Fig:separated_paths} can also be identified. In addition, the trend of the received power of MPCs in Fig.~\ref{Fig:separated_paths} is used to predict the death of the MPCs in Section~\ref{Section:predict_death}. The trend of the other five channel parameters can be shown in a similar way.

\begin{figure}[!h]
	\centering\vspace{-2mm}
	\includegraphics[width=0.85\columnwidth, scale=0.5]{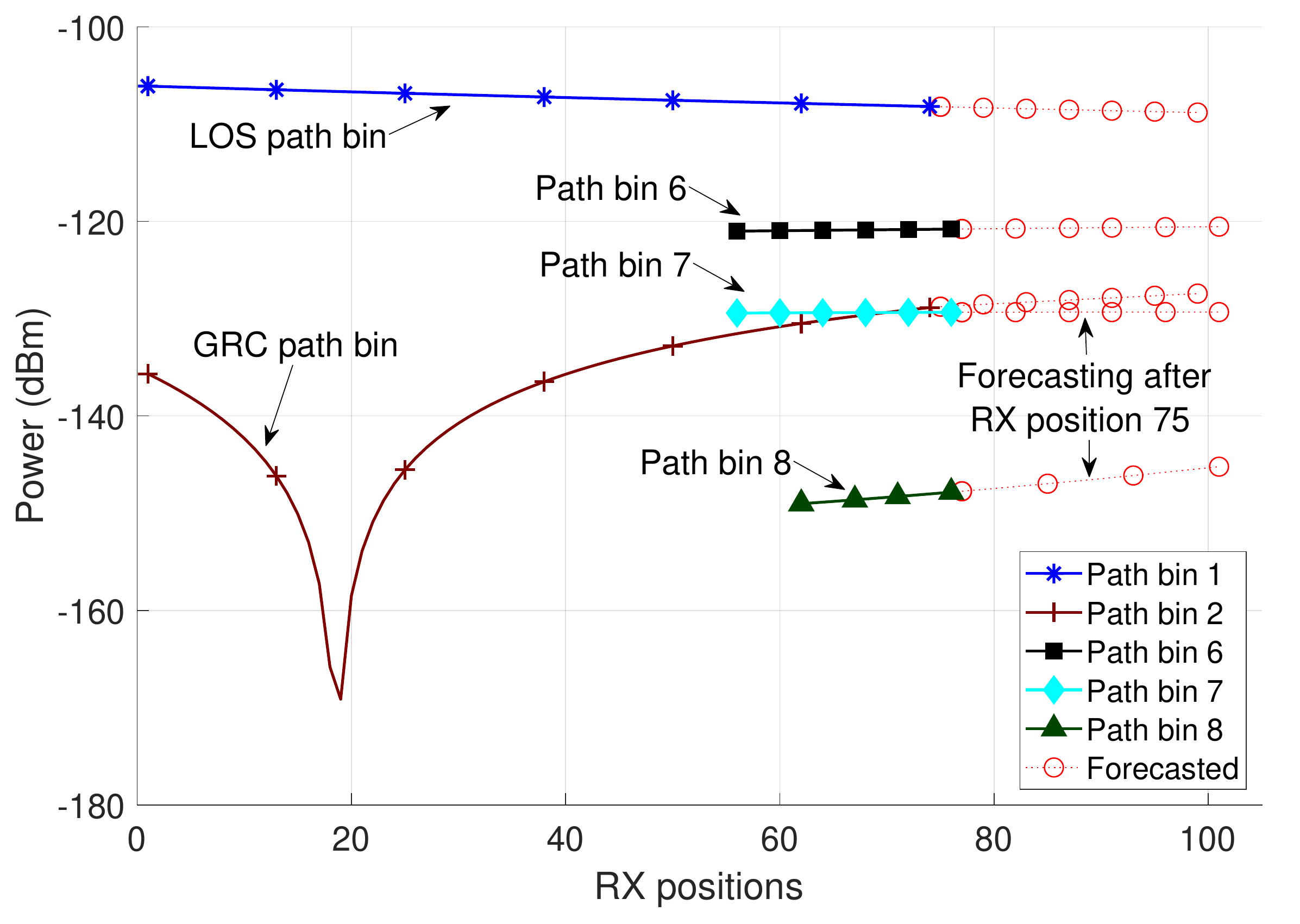}\vspace{-3mm}
	\caption{Forecasting received power of MPCs in path bins during the blockage.}\label{Fig:separated_paths_osbstructed_forecasted}\vspace{-1mm}
\end{figure}

A blockage is now considered at RX position 75. All the paths from RX position 75 onwards are removed. The channel parameters of the MPCs in path bins after RX position 75 are forecasted using Algorithm~\ref{Alg:forecasting}, that uses a Matlab based fourth-order AR model and forecast function. The forecasted power of the MPCs in path bins is shown in Fig.~\ref{Fig:separated_paths_osbstructed_forecasted}. Comparing Fig.~\ref{Fig:separated_paths_osbstructed_forecasted} and Fig.~\ref{Fig:separated_paths}, we observe that the forecasted and actual values are close and the mean square error~(MSE) between the two is given in Table~\ref{Table:comparison}. The other channel parameters are also forecasted during blockage in a similar way.

\subsection{Prediction of Death of MPCs} \label{Section:predict_death} 
From Fig.~\ref{Fig:separated_paths}, we can observe the birth and death of the MPCs and corresponding path bins. We can also predict the death of a path bin~(containing a resolvable MPC sequence) using Fig.~\ref{Fig:Markov}. Let us take the LOS as the reference and denote the number of RX positions where the $l^{\rm th}$ path bin~(from Algorithm~\ref{Alg:Algorithm}) is not empty as $N_l$. Then, the average distance between the channel parameters of the $l^{\rm th}$ path bin and the LOS over $N_l$ RX positions is represented as $d_l = \sum_{i=j}^{N_l+j}\frac{d(i,l,i,1)}{N_l}$, using (\ref{Eq:distance2}). 
The distance $d_l$ can be used to predict the death of the $l^{\rm th}$ path bin along the RX positions, where $l=2,3,...,8$, for the seven path bins, excluding the LOS. The greater the distance $d_l$, the greater the likelihood of the death of the $l^{\rm th}$ path bin. For the seven path bins in Fig.~\ref{Fig:Path_bins}, $d_l = [0.68, 5.07, 4.29, 4.45, 3.85, 4.01, 4.18]$. 
 If a distance threshold of $4.20$ is used, then path bins $3$, $4$, and $5$ with average distances greater than the threshold are predicted to die. 
This can also be observed in Fig.~\ref{Fig:separated_paths}, where the received power of path bins is shown. A general observation can be drawn that the farther the channel parameters of a path bin from the channel parameters of the LOS over the RX positions, the greater the likelihood that the path bin will die.

A comparison of our approach with AR and long short-term memory RNN based prediction is provided in Table~\ref{Table:comparison}. 
Our approach has higher accuracy compared to the other two methods due to accurate tracking/clustering of MPCs into respective path bins, which comes at the expense of increased computational complexity. The computational complexity for our approach consists mainly of distance calculation provided in Algorithm~\ref{Alg:Algorithm}, and forecasting using AR model given in Algorithm~\ref{Alg:forecasting}. The computational complexity of RNN in Table~\ref{Table:comparison} is obtained from \cite{NN4}, where $n_{\rm c}$ represents the number of memory cells. Overall, our approach can help to identify spur MPCs~\cite{spur}, estimate the number of scatterers in the environment, observe the birth, death, survival, resurrection, and trend of evolution of MPCs along the UAV trajectory in practical environments. 

\begin{table}[h]\vspace{-2mm}
	\begin{center}
      \footnotesize
		\caption{Comparison of our approach with popular prediction methods.} \label{Table:comparison}\vspace{-3mm}
\begin{tabular}{@{}|P{1.2cm}|P{3.75cm}|P{0.6cm}|P{0.6cm}|P{0.65cm}|@{}}
 \hline
Predictor&Computational complexity&MSE [dB]&Stages&MAPE
\\
\hline
This work&$O\Big[\big(6(N-1)M^2\big)^2 +6nNM\Big]$&9.56 &2&2.25\%
\\
\hline
AR model&$6nNM$&17.4&1&38.5\%
\\
\hline
RNN&O\Big[4$n_{\rm c}^2$+30$NMn_{\rm c}$+3$n_{\rm c}$\Big]&16&2&31.6\%
\\
\hline
\end{tabular}
		\end{center}\vspace{-4mm}
		\end{table} 

\vspace{-2mm}

\section{Concluding Remarks} \label{Section:Conclusions}
In this work, we have introduced a novel channel prediction method during GA link blockages. This method uses a ED based algorithm for arranging the MPCs into path bins. The arrangement places the MPCs with similar channel characteristics in individual path bins. The channel parameters of the MPCs in path bins are then forecasted during a link blockage with high accuracy. A distance-based prediction for the death of a MPC sequence is also provided.

\ifCLASSOPTIONcaptionsoff
  \newpage
\fi

\bibliographystyle{IEEEtran}



\end{document}